\documentstyle[11pt]{article}
\input tcilatex

\begin{document}

\title{{\bf The power-law and the logarithmic potentials}}
\author{{\bf Hakan \d{C}\.{I}FTC\.{I}} \thanks{%
e-mail:hciftci@gazi.edu.tr}, {\bf Engin ATE\d{S}ER and H\"{u}seyin KORU} 
\thanks{%
e-mail:hkoru@gazi.edu.tr} \\
%EndAName
{\it Gazi University, Faculty of Art and Science, 06500 Ankara, TURKEY}}
\maketitle

\begin{abstract}
In this study, we show that the energy eigenvalues and the eigenfunctions of
the Schrodinger equation for the power-law and the logarithmic potential can
be easily obtained by using variation technique for special type wave
functions. The results are in very good agreement with exact numerical
results.
\end{abstract}

\baselineskip 0.8 cm

\section{Introduction}

A large number of important physical problems require solving the
Schrodinger equation for spherical symmetric potential to determine the
energy eigenvalues and the eigenfunctions. It is known that for very limited
potentials, Schrodinger equation is exactly solvable. In general one has to
resort to numerical techniques or approximation schemes, Most popular
approximation methods like $1/N$ expansion, WKB method, perturbation theory
are widely used for this purpose. But some of these methods have-draw backs
in application. Although some methods give simple relations for the
eigenvalues, they give very complicate relations for the eigenfunction. The
aim of present work is to give a simple way for finding both eigenvalues and
the eigenfunctions of Schrodinger and Schrodinger-like equations for
power-law and logarithmic potentials, which are very important in particle
physics[1-10]. This paper is organized as follows. In the first section, the
eigenvalues are obtained for the $sgn(v)Ar^{\nu }(\nu >-2)$ type potential
by using variation technique for the special form of eigenfunction. In
section 2, the same technique is applied to the logarithmic potential. In
the last section, we give some concluding remarks.

\section{The solution of the Schrodinger equation for the power-law
potentials}

The radial part of Schrodinger equation for $sgn(v)Ar^{\nu }(\nu >-2)$ type
potential is written as 
\begin{equation}
\left[ -\frac{d^{2}}{dr^{2}}+sgn(v)Ar^{\nu }+\frac{l(l+1)}{r^{2}}\right]
g(r)=Eg(r)
\end{equation}
Using substitution $r=(\frac{1}{A})^{\frac{1}{\nu +2}}\rho $, Eq. (1) can be
written in the following form 
\begin{equation}
\left[ -\frac{d^{2}}{d\rho ^{2}}+sgn(v)\rho ^{\nu }+\frac{l(l+1)}{\rho ^{2}}%
\right] g(\rho )=\epsilon g(\rho )
\end{equation}
where $\epsilon =E.(\frac{1}{A})^{\frac{2}{\nu +2}}.$ It is well known that
Eq. (2) have exact analytical solutions for $\nu =-1$ and $\nu =2$ and the
forms of the solutions are:\newline
for $\nu =-1$ (Coulomb potential) 
\begin{equation}
g(\rho )\approx \rho ^{l+1}Exp\left( -x\rho \right) L_{n}^{2l+1}(2x\rho ),
\end{equation}
for $\nu =2$ (Harmonic potential) 
\begin{equation}
g(\rho )\approx \rho ^{l+1}Exp\left( -\left( x\rho \right) ^{2}\right)
L_{n}^{\frac{2l+1}{2}}(2(x\rho )^{2}).
\end{equation}
where $L$ is Laguerre polynomials. From the similarity of these solutions,
the following solution can be proposed for arbitrary $\nu $ 
\begin{equation}
g(\rho )\approx \rho ^{l+1}Exp\left( -\left( x\rho \right) ^{d}\right)
L_{n}^{\frac{2l+1}{d}}(2(x\rho )^{d})
\end{equation}
where $x$ and $d$ variation parameters and they can be obtained by
minimizing of $\epsilon $ in Eq. (2) with respect to these parameters. So; 
\begin{equation}
\frac{\partial \varepsilon }{\partial x}=0,
\end{equation}
\begin{equation}
\frac{\partial \varepsilon }{\partial d}=0.
\end{equation}
Using Eqs. (2) and (5), for $\epsilon $, we get 
\begin{equation}
\epsilon _{nl}(x\text{,}d)=cx^{2}+\frac{b}{x^{v}}
\end{equation}
and from Eq. (6), we have $x=\left( \frac{b\nu }{2c}\right) ^{\frac{1}{\nu +2%
}},$thus, we have found \ $\epsilon _{nl}$ as the following form\ \ \ \ \ \
\ \ \ \ \ \ \ \ 

\begin{equation}
\epsilon _{nl}(d)=\left( v+2\right) \left( \frac{c}{\nu }\right) ^{\frac{\nu 
}{v+2}}\left( \frac{b}{2}\right) ^{\frac{2}{v+2}},
\end{equation}
where;

\ \ \ \ \ \ \ \ \ \ \ \ \ \ \ \ \ \ 
\begin{equation}
c=2^{\frac{2-2d}{d}}\frac{\sum_{k=0}^{n}\sum_{m=0}^{n}a_{k}a_{m}s\Gamma (k+m+%
\frac{2l+1}{d})}{\sum_{k=0}^{n}\sum_{m=0}^{n}a_{k}a_{m}\Gamma (k+m+\frac{2l+3%
}{d})},
\end{equation}

\bigskip

\begin{equation}
b=sgn(v)2^{\frac{-\nu }{d}}\frac{\sum_{k=0}^{n}\sum_{m=0}^{n}a_{k}a_{m}%
\Gamma (k+m+\frac{2l+\nu +3}{d})}{\sum_{k=0}^{n}\sum_{m=0}^{n}a_{k}a_{m}%
\Gamma (k+m+\frac{2l+3}{d})},
\end{equation}

\bigskip

\begin{equation}
s=(2l+1)(2l+d+1)+(k+m-(k-m)^{2})d^{2},
\end{equation}
where $a_{j}$ are the coefficients of the generalized Laguerre polynomials.
The parameter $d$ can be obtained from Eq. (7), but unfortunately this
equation can not be solved analytically. However, for $n=0$, $l=0$ case, the
value of $d$ which minimized of $\epsilon $ can be found easily. In Fig. 1,
we present the dependence $d$ on $\nu $. Behavior of the curve is similar to 
$\sqrt{v+2}$ function. But, this function does not fit exactly to the curve.
Therefore, a correction factor is necessary. According to the our
assumption, the correction factor must be equal to $1$ at $v=-1$ and $v=2$.
Thus, we chose the correction factor in the following form; 
\begin{equation}
w=(1+tp)^{h}
\end{equation}
where $p=\frac{(\nu +1)(2-\nu )}{a_{1}\nu ^{2}+a_{2}\nu +a_{3}}$. So, we
write

\bigskip 
\begin{equation}
d=\sqrt{v+2}w
\end{equation}
\ \ Fitting this equation to the curve in Fig. 1, $t$, $a_{1}$, $a_{2}$, $%
a_{3}$ and $h$ constants are obtained as, $0.2075,$ $0.1381$, $1..05$, $%
2.484 $ and $0.08104$ respectively$.$ So, when any values of $v$ is given, $%
d $ values corresponding to $v$ can be easily calculated by using Eq. (14).
In Tables 1 and 2, we present results of our calculations for the
eigenvalues at different $\nu .$ For comparison in these tables, we also
present existing numerical solutions in literature .In Table 2, the
calculations done using eq. (9) are multiplied by $2^{\frac{-v}{v+2}}$factor
because of consistency with results of ref[11].

\bigskip

When we study on linear potentials, Eq. (2) has the following form for S
states

\bigskip

\begin{equation}
\left[ -\frac{d^{2}}{dz^{2}}+z\right] g(z)=0
\end{equation}

\bigskip

\bigskip where, $z=\rho -\epsilon $. The equation given above is known as
Airy equation and has exact solution in terms of Airy functions given as the
following form

\bigskip 
\begin{equation}
g_{n}^{\text{exact}}(\rho )=Ai(\rho -\epsilon _{n})
\end{equation}

Where $\epsilon _{n}$ are the zeros of the Airy function and as given in
Table 3. For S states, our prediction for eigenfunctions has the following
form from Eq. (5)

\bigskip

\begin{equation}
g_{n}^{\text{our}}(\rho )\approx \rho ^{l}Exp\left( -\left( x\rho \right)
^{d}\right) L_{n}^{\frac{1}{d}}(2(x\rho )^{d})
\end{equation}

In Figures 2 and 3, our wave function and the exact wave function are
presented together for $n=0$, $l=0$ and $n=4$, $l=0$ states respectively. We
see that the agreement between two solutions is excellent. In addition to
the wave functions, in Table 3, eigenvalues of the linear potential for S
states ($n=0$, $1$,$2$,.. and $l=0)$ are given and compared with the exact
results which are well-known zeros of the Airy function. Similarly, in Table
4, eigenvalues of $r^{0.5}$ are calculated and compared with exact numerical
results. The calculated results are in good agreement with exact numerical
results and better than given in ref[17].

\section{The logarithmic potential case}

Let us consider $V(r)=\log (r)$ potential which is very important in
particle physics. The radial part of Schrodinger equation for logarithmic
potential is written as 
\begin{equation}
\left[ -\frac{d^{2}}{dr^{2}}+\log (r)+\frac{l(l+1)}{r^{2}}\right] g(r)=Eg(r).
\end{equation}
The function $\log (r)$ at $\nu \cong 0$ can be written in following form 
\begin{equation}
\log (r)\cong \frac{1}{\nu }\left[ r^{\nu }-1\right] \text{.}
\end{equation}
Substitute Eq. (15) into Eq. (14) and then apply the method presented in
previous section to Eq. (14) for $d$ we found the value $1.43203$ at $\nu
=0.00001$ (see Eq. (13)). Thus, the eigenvalues of Eq. (14) are written as 
\begin{equation}
E_{nl}=\frac{\epsilon _{nl}}{\nu ^{\frac{2}{\nu +2}}}-\frac{1}{\nu }
\end{equation}
where $\epsilon _{nl}$ is given as in Eq. (8). The calculated results for $%
\epsilon _{nl}$ are presented in Table 4. In this table we also give the
results of numerical solutions. The eigenfunctions of logarithmic potential
can be written from Eq. (5) as 
\begin{equation}
g(\rho )\approx \rho ^{l+1}Exp\left( -\left( x\rho \right) ^{d}\right)
L_{n}^{\frac{2l+1}{d}}(2(x\rho )^{d})
\end{equation}
where, $\rho =\frac{r}{\nu ^{\frac{1}{\nu +2}}}.$ It is obvious that the
eigenvalues obtained for the logarithmic potential are in good agreement
with the results obtained from numerical solution. Also, in order to show
the validity of the wavefunctions given in Eq. (17) obtained for the
logarithmic potential, first of all, some energy levels are chosen. Their
wavefunctions are obtained by Numerov's Method. Finally, these results are
given in Figures 4 and 5 together with predictions.

\section{Conclusion}

In this paper, we have calculated eigenvalues and eigenfunctions of
power-law and logarithmic potentials by using variational techniques for
special type wave functions. Our results are in good agreement with existing
exact numerical ones. However, for the higher values of $\nu $, there is
some difference between two approaches. However this differences not larger
than that one coming from 1/N expansion. Moreover, we obtained that present
method predicts not only the eigenvalues as well as the eigenfunctions of
given potentials. The correspondence of wavefunctions for lower energy
levels is very good agreement with numerical results as expected the except
case $n$ and $l$ are large. The method used in this study present an easy
way of calculating both eigenfunctions and eigenvalues of power-law and
logarithmic potentials. In spite of its simple structure, the method is very
practical and the results of the method are in good agreement with exact
results.\ 

\section{Acknowledgment}

The authors thank Prof. Dr. T. M. Aliev and Prof. Dr. M. \d Sim\d sek for
useful discussions.

\section{References}

\begin{enumerate}
\item  E. Magyari, $Phys.Lett.\ B$ {\bf 95} (1980), 295.

\item  C. Quigg and J. L. Rosner, $Phys.Lett.\ B$ {\bf 71} (1977), 153.

\item  S. N. Jena and D. P. Rath, $Phys.Rev.\ D$ {\bf 34} (1986), 196.

\item  N. Barik, S. N. Jena, D. P. Rath, $Phys.Rev.\ D$ {\bf 41} (1990),
1568.

\item  N. Barik, S. N. Jena, $Phys.Lett.\ B$ {\bf 97} (1980), 265.

\item  S. N. Jena, P. Panda and T. C. Tripathy, $Phys.Rev.\ D$ {\bf 63},
(2000) 014011-1.

\item  S. N. Jena, P. Panda and T. C. Tripathy, $J.Phys.G$ {\bf 27} (2001),
227.

\item  H. Akcay and H. Ciftci, $J.Phys.\ G$ {\bf 22} (1996), 455.

\item  H. Ciftci and H. Koru, $Int.J.Mod.Phys.\ E$ {\bf 9} (2000), 407.

\item  A. Martin, $Phys.Lett.\ B$ {\bf 93} (1980),338; {\bf 100} (1981),511.

\item  T. Imbo, A. Pagnamenta and U. Sukhatme, $Phys.Rev.\ D$ {\bf 29}
(1984), 1669.

\item  C. Quigg and J. L. Rosner, $Phys.Rep.$ {\bf 56} (1979),206.

\item  J. Richardson and R. Blankenbecler, $Phys.Rev.\ D$ {\bf 19} (1979),
496.

\item  M. Dumont-LePage, N. Gani, J. Gazeau and A. Ronveaux, $J.\ Phys.\ A$ 
{\bf 13} (1980),1243.

\item  C. Bender and S. Orzag, Advanced Mathematical Methods for Scientists
and Engineers, McGraw-Hill , New York, 1978.

\item  R. N. Faustov, V. O. Galkin, A. V. Tatarintsev and A. S. Vshivtsev, $%
Int.J.Mod.Phys.\ A$ {\bf 15} (2000), 209.

\item  R. L. Hall, $J.Phys.G.${\bf 26 }(2000),981.
\end{enumerate}

\newpage\ 

Table.1. Comparison of this study results for the ground state ($n=0,$ $l=0)$
of various power-law potentials.

\begin{tabular}{|c|c|c|c|c|c|}
\hline\hline
$V(r)$ & This Work & Numerical[11] & $V(r)$ & This work & Numerical[11] \\ 
\hline
$-r^{-1.5}$ & -0.29703 & -0.29609 & $r^2$ & 3 & 3 \\ \hline
$-r^{-1.25}$ & -0.22027 & -0.22029 & $r^3$ & 3.45110 & 3.45056 \\ \hline
$-r^{-1}$ & -0.25 & -0.25 & $r^4$ & 3.80241 & 3.79967 \\ \hline
$r^0$ & 1 & 1 & $r^5$ & 4.09626 & 4.33801 \\ \hline
$r^{0.15}$ & 1.32798 & 1.32795 & $r^6$ & 4.35243 & 4.54690 \\ \hline
$r^{0.5}$ & 1.83352 & 1.83339 & $r^7$ & 4.58158 & 4.71772 \\ \hline
$r^{0.75}$ & 2.10829 & 2.10814 & $r^8$ & 4.79013 & 4.92220 \\ \hline
$r^{1.5}$ & 2.70816 & 2.70809 & $r^{10}$ & 5.16092 & - \\ \hline
\end{tabular}
\newpage\ Table.2 Eigenvalues of $-2^{1.7}r^{-0.2}$ and $-2^{0.8}r^{-0.8}$
for different $nl$

\begin{tabular}{|c|c|c|c|c|c|c|c|}
\hline\hline
$n$ & $l$ & This work & numerical[11-15] & $n$ & $l$ & This work & 
numerical[11-15] \\ \hline
0 & 0 & -2.6859 & -2.686 & 0 & 0 & -1.2186 & -1.218 \\ \hline
1 & 0 & -2.2530 & -2.253 & 1 & 0 & -0.4622 & -0.462 \\ \hline
2 & 0 & -2.0440 & -2.044 & 2 & 0 & -0.2648 & -0.265 \\ \hline
0 & 1 & -2.3449 & -2.345 & 0 & 1 & -0.5004 & -0.500 \\ \hline
1 & 1 & -2.1006 & -2.101 & 1 & 1 & -0.2806 & -0.281 \\ \hline
2 & 1 & -1.9504 & -1.951 & 2 & 1 & -0.1873 & -0.187 \\ \hline
0 & 2 & -2.1562 & -2.156 & 0 & 2 & -0.2947 & -0.295 \\ \hline
1 & 2 & -1.9900 & -1.990 & 1 & 2 & -0.1949 & -0.195 \\ \hline
2 & 2 & -1.8749 & -1.875 & 2 & 2 & -0.1420 & -0.142 \\ \hline
0 & 3 & -2.0291 & -2.029 & 0 & 3 & -0.2019 & -0.202 \\ \hline
1 & 3 & -1.9049 & -1.905 & 1 & 3 & -0.1463 & -0.146 \\ \hline
2 & 3 & -1.8124 & - & 2 & 3 & -0.1128 & - \\ \hline
\end{tabular}

\newpage\ 

\begin{center}
Table.3 Eigenvalues of linear potential for different $n$ and $l=0.$\\[0pt]

\begin{tabular}{|c|c|c|c|}
\hline\hline
$n$ & $l$ & This work & numerical[16] \\ \hline
0 & 0 & 2.33825 & 2.33810 \\ \hline
1 & 0 & 4.08918 & 4.08795 \\ \hline
2 & 0 & 5.52132 & 5.52056 \\ \hline
3 & 0 & 6.78614 & 6.78671 \\ \hline
4 & 0 & 7.94189 & 7.94413 \\ \hline
5 & 0 & 9.01859 & 9.02265 \\ \hline
\end{tabular}

\bigskip

\bigskip \newpage\ 

Table.4 Eigenvalues of $r^{0.5}$ for different $nl$ together with exact
values and other researcher results, with percentage errors

\begin{tabular}{|l|l|l|l|l|l|}
\hline\hline
$n$ & $l$ & This work & Numerical[17] & Ref[17] & \% \\ \hline
0 & 0 & 1.83352 & 1.83339 & 1.83375 & 0.007 \\ \hline
1 & 0 & 2.55152 & 2.55065 & 2.55142 & 0.03 \\ \hline
2 & 0 & 3.05177 & 3.05118 & 3.05224 & 0.019 \\ \hline
3 & 0 & 3.45197 & 3.45213 & 3.45341 & 0.005 \\ \hline
4 & 0 & 3.79233 & 3.79336 & 3.79482 & 0.027 \\ \hline
0 & 1 & 2.30056 & 2.30050 & 2.30073 & 0.003 \\ \hline
1 & 1 & 2.85473 & 2.85434 & 2.85486 & 0.014 \\ \hline
2 & 1 & 3.28666 & 3.28583 & 3.28659 & 0.025 \\ \hline
3 & 1 & 3.64838 & 3.64739 & 3.64835 & 0.027 \\ \hline
4 & 1 & 3.96361 & 3.96268 & 3.96382 & 0.023 \\ \hline
0 & 2 & 2.65760 & 2.65756 & 2.65775 & 0.002 \\ \hline
1 & 2 & 3.12048 & 3.12033 & 3.12077 & 0.005 \\ \hline
2 & 2 & 3.50296 & 3.50245 & 3.50309 & 0.015 \\ \hline
3 & 2 & 3.83338 & 3.83254 & 3.83336 & 0.022 \\ \hline
4 & 2 & 4.12686 & 4.12581 & 4.12678 & 0.025 \\ \hline
0 & 3 & 2.95448 & 2.95445 & 2.95461 & 0.001 \\ \hline
1 & 3 & 3.35764 & 3.35759 & 3.35798 & 0.001 \\ \hline
2 & 3 & 3.70299 & 3.70270 & 3.70327 & 0.008 \\ \hline
3 & 3 & 4.00796 & 4.00737 & 4.00810 & 0.015 \\ \hline
4 & 3 & 4.28282 & 4.28196 & 4.28283 & 0.020 \\ \hline
0 & 4 & 3.21236 & 3.21233 & 3.21247 & 0.001 \\ \hline
1 & 4 & 3.57275 & 3.57275 & 3.57310 & 0.000 \\ \hline
2 & 4 & 3.88913 & 3.88898 & 3.88950 & 0.004 \\ \hline
3 & 4 & 4.17308 & 4.17268 & 4.17335 & 0.010 \\ \hline
4 & 4 & 4.43196 & 4.46131 & 4.43164 & 0.015 \\ \hline
\end{tabular}
\end{center}

\newpage\ Table.5 Eigenvalues of logarithmic Potential for different $nl$

\begin{tabular}{|c|c|c|c|c|c|c|c|}
\hline\hline
$n$ & $l$ & This work & numerical[11-15] & $n$ & $l$ & This work & 
numerical[11-15] \\ \hline
0 & 0 & 1.0445 & 1.0443 & 3 & 0 & 2.5957 & 2.5957 \\ \hline
0 & 1 & 1.6412 & 1.6430 & 3 & 1 & 2.7465 & 2.7440 \\ \hline
0 & 2 & 2.0134 & 2.0150 & 3 & 2 & 2.8801 & 2.8800 \\ \hline
0 & 3 & 2.2842 & 2.2860 & 3 & 3 & 2.9996 & 2.9990 \\ \hline
1 & 0 & 1.8485 & 1.8474 & 3 & 4 & 3.1071 & 3.1070 \\ \hline
1 & 1 & 2.1513 & 2.1510 & 4 & 0 & 2.8293 & 2.8299 \\ \hline
1 & 2 & 2.3875 & 2.3880 & 4 & 1 & 2.9498 & 2.9480 \\ \hline
1 & 3 & 2.5798 & 2.5810 & 4 & 2 & 3.0592 & 3.0600 \\ \hline
2 & 0 & 2.2903 & 2.2897 & 4 & 3 & 3.1592 & 3.1590 \\ \hline
2 & 1 & 2.4917 & 2.4910 & 4 & 4 & 3.2512 & 3.2510 \\ \hline
2 & 2 & 2.6629 & 2.6630 & 6 & 0 & 3.1770 & 3.1791 \\ \hline
2 & 3 & 2.8106 & - & 10 & 0 & 3.6411 & 3.6427 \\ \hline
\end{tabular}

\newpage\ 

\section*{Figure captions}

Figure 1. The dependence of $d$ on $\nu $.\newline

Figure 2. Comparison of our wavefunction and Airy function for linear
potential ($n=0$, $l=0$),Dot lines represent Airy function,

Figure 3. Comparison of our wavefunction and Airy function for linear
potential ($n=4$, $l=0$),Dot lines represent Airy function,

Figure 4. Comparison of our wavefunction and corresponding numerical
wavefunction for logarithmic potential \ \ \ \ \ 

($n=0$, $l=0$)

Figure 5. Comparison of our wavefunction and corresponding numerical
wavefunction for logarithmic potential \ \ \ \ \ \ \ \ \ \ \ \ \ \ \ \ 

($n=4$, $l=4$)

\end{document}